\documentclass{article}
\usepackage{dcase2017,amsmath,graphicx,url,times,booktabs, tabularx}

\title{SURREY-CVSSP SYSTEM FOR DCASE2017 CHALLENGE TASK4}

\twoauthors
  {Yong Xu\textsuperscript{*},  Qiuqiang Kong\textsuperscript{*}\thanks{The first two authors have equal contribution.}}
    {University of Surrey\\
Centre for Vision, Speech and Signal Processing\\
     Guildford, GU2 7XH, UK \\
     \{q.kong, yong.xu\}@surrey.ac.uk}
  {Wenwu Wang, Mark D. Plumbley\thanks{This work was supported by the Engineering and Physical Sciences Research Council (EPSRC) of the UK under the grant EP/N014111/1. Qiuqiang Kong is partially supported by China scholarship council (CSC).}}
    {  University of Surrey \\
     Centre for Vision, Speech and Signal Processing\\
     Guildford, GU2 7XH, UK \\
     \{w.wang, m.plumbley\}@surrey.ac.uk}

\begin{document}

\ninept
\maketitle

\begin{sloppy}

\begin{abstract}
In this technical report, we present a bunch of methods for the task 4 of Detection and Classification of Acoustic Scenes and Events 2017 (DCASE2017) challenge. This task evaluates systems for the large-scale detection of sound events using weakly labeled training data. The data are YouTube video excerpts focusing on transportation and warnings due to their industry applications. There are two tasks, audio tagging and sound event detection from weakly labeled data. Convolutional neural network (CNN) and gated recurrent unit (GRU) based recurrent neural network (RNN) are adopted as our basic framework. We proposed a learnable gating activation function for selecting informative local features. Attention-based scheme is used for localizing the specific events in a weakly-supervised mode. A new batch-level balancing strategy is also proposed to tackle the data unbalancing problem. Fusion of posteriors from different systems are found effective to improve the performance. In a summary, we get 61\% F-value for the audio tagging subtask and 0.73 error rate (ER) for the sound event detection subtask on the development set. While the official multilayer perceptron (MLP) based baseline just obtained 13.1\% F-value for the audio tagging and 1.02 for the sound event detection. Finally, we ranked first in the audio tagging sub-task on the evaluation set. We also ranked 2nd as a team in the sound event detection sub-task.
\end{abstract}


\section{Introduction}
\label{sec:intro}

DCASE challenge has been organized for three rounds since 2013 \cite{giannoulis2013detection, stowell2015detection, dcase2016}. It urges lots of academic and industrial effort to make analysis and utilizing of the environmental audio. This series of challenges consists three parts, including acoustic scene classification \cite{stowell2015detection}, sound event detection \cite{stowell2015detection} and audio tagging \cite{dcase_t4, Foster2016}.

This technical report summaries the methods we use on the task 4 of DCASE2017, the large-scale weakly supervised sound event detection. Task 4 dataset of DCASE 2017 is a subset of Google Audioset \cite{audioset}, consisting recordings of a variety of cars. Task 4 consists of two subtasks, audio tagging and sound event detection. Both of the two subtasks are based on weakly labelled audio data. For audio tagging, lots of deep learning based methods have been developed \cite{yong2017ijcnn, xu2017trans, xu2016fully, Cakir2016}. As for weakly supervised sound event detection, multiple instance learning \cite{kumar2016audio}, joint detection and classification (JDC) \cite{qq2017icassp} and attention-based methods \cite{yong2017interspeech} have been proposed. These weakly labelled learning methods such as the JDC and attention methods have not been well evaluated before the task 4 of DCASE 2017 appear. 

In this report, we will present our methods both for audio tagging and weakly-supervised sound event detection.

\section{Proposed methods for audio tagging}
\label{sec:format}

We present a bunch of methods to solve this problem, including data balancing, gated activation function, auto threshold and posterior fusion. 

\subsection{Data balancing}
There are 17 classes in the whole dataset\footnote{http://www.cs.tut.fi/sgn/arg/dcase2017/challenge/task-large-scale-sound-event-detection}. However the number of samples of each class is not balanced. For example, the sample number of ``car'' is 25744 while ``Car alarm'' only has 273 samples. Note that the neural network model is trained in mini batch with stochastic gradient decent. The batch size is always fixed, for example, 128. The 128 samples will be randomly selected from the whole dataset, which means the class with less samples will have low chance to be selected. The neural network is easily biased by the audio samples with large number. Hence, we redesign the sample selection scheme for generating each batch. We almost follow the distribution ratio of each class in the whole dataset, but we also make sure that there is at least one sample for the low probability event in each batch.

\subsection{Learnable gated activation function}
Sigmoid or ReLU might be used for the activation function in the neural network to introduce the non-linear characteristics. However, in this work, we use a learnable gated activation function \cite{dauphin2016language}.

\begin{equation}
\textbf{Y1}=linear(\textbf{X} \ast \mathbf{W} + b)
\end{equation}
\begin{equation}
\textbf{Y2}=sigmoid(\textbf{X} \ast \mathbf{V} + c)
\end{equation}
\begin{equation}
\textbf{Y}=\textbf{Y1}\otimes{\textbf{Y2}}
\end{equation}
Where $\textbf{X}$ is the input feature, $\textbf{Y1}$ is the linear output with the parameters $\textbf{W1}$. $\textbf{Y2}$ is the learned gating matrix through parameters $\textbf{W2}$. Finally, the output $\textbf{Y}$ is obtained by the element-wise multiplication of $\textbf{Y1}$ and $\textbf{Y2}$. This gating activation function can be regarded as a local attention scheme.

\subsection{Auto threshold}
To decide the existence or presence of an acoustic event, we should choose a hard threshold. Basically, a fixed value will be chosen for each class. However, each class should have a specific threshold. We tuned these thresholds for each class on the development set, and applied them onto the private set.

\subsection{System fusion}
Fusion is always important to get a robust result. We use two-level fusion. One is to fuse the iterations among the same model. The 2nd level is to fuse the posteriors of different models.

\subsection{Whole framework}
We treat one CNN layer followed by one batch normalization layer and the learned gating layer as a block. Two this kind of blocks and one max-pooling layer are a unit. Four units were concatenated together to form the whole system. The final output layer is the weighted mean along the time axis to get the 17 output nodes.

\section{Proposed methods for weakly supervised sound event detection}
The basic framework is similar to the system for the audio tagging. However, we use an additional bidirectional GRU-RNN after the CNN layers. We also did not execute any pooling operations through the time axis to keep the location information.

\section{Experiments and results}
\subsection{Experimental setup}
The task employs a subset of Google Audioset \cite{audioset}. AudioSet consists of an expanding ontology of 632 sound event classes and a collection of 2 million human-labeled 10-second sound clips (less than 21\% are shorter than 10-seconds) drawn from 2 million YouTube videos. The ontology is specified as a hierarchical graph of event categories, covering a wide range of human and animal sounds, musical instruments and genres, and common everyday environmental sounds. The subset consists of 17 sound events divided into two categories: ``Warning'' and ``Vehicle''. The list below shows each class and next to it the approximate number of 10-second clips. Note that each clip may correspond to more than one sound event.

We extracted spectrogram, log-mel filter banks, Mel-frequency cepstrum coefficients (MFCC). Each utterance has 240 frames.

\subsection{Results}
In this section, the audio tagging results and the weakly supervised sound event detection results will be given.
\subsubsection{Audio tagging}
Table \ref{tab:at} presents the F1, Precision and Recall comparisons for the \textbf{audio tagging} sub-task on the development set and the evaluation set. We finally ranked first in the audio tagging subtasks among 29 systems.
\begin{table}[h]  
	\centering
	\caption{F1, Precision and Recall comparisons for the \textbf{audio tagging} sub-task on the development the {evaluation sets}.}
	\begin{tabular}{lccc}
		\hline
		\textbf{Dev-set} & \textbf{F1} & \textbf{Precision} & \textbf{Recall} \\
		\hline
		DCASE2017 Baseline \cite{mesaros2017dcase} & 10.9  & 7.8   & 17.5 \\
		\hline
		CRNN-logMel-noBatchBal & 42.0  & 47.1  & 38.0 \\
		\hline
		CRNN-logMel & 52.8  & 49.9  & 56.1 \\
		\hline
		Gated-CRNN-logMel (i) & 56.7  & 53.8  & \textbf{60.1} \\
		\hline
		Gated-CRNN-MFCC (ii) & 52.1  & 51.7  & 52.5 \\
		\hline
		Fusion (i+ii) & \textbf{57.7} & \textbf{56.5} & 58.9 \\
		\hline
		\hline
		\textbf{Eval-set} & \textbf{F1} & \textbf{Precision} & \textbf{Recall} \\
		\hline
		DCASE2017 Baseline & 18.2  & 15.0  & 23.1 \\
		\hline
		CNN-ensemble & 52.6  & \textbf{69.7} & 42.3 \\
		\hline
		Frame-CNN & 49.0  & 53.8  & 45.0 \\
		\hline
		Our gated-CRNN-logMel & 54.2  & 58.9  & 50.2 \\
		\hline
		Our fusion system & \textbf{55.6} & 61.4  & \textbf{50.8} \\
		\hline
	\end{tabular}%
	\label{tab:at}
\end{table}

\subsubsection{Weakly supervised sound event detection (SED)}
The results of F1 and Error rate comparisons on the development set and the evaluation set for the 2nd SED task are given in Table \ref{tab:sed}. With the fusion system, we rank 2nd as a team in the sound event detection sub-task.
\begin{table}[h]  
	\centering
	\caption{The results of F1 and Error rate comparisons on the development set and the evaluation set for the \textbf{sound event detection} sub-task among several methods across the 17 audio event tags.}
	\begin{tabular}{lcc}
		\hline
		\textbf{Dev-set} & \textbf{F1} & \textbf{Error rate} \\
		\hline
		DCASE2017 baseline & 13.8  & 1.02 \\
		\hline
		Gated-CRNN-logMel & 47.20 & 0.76 \\
		\hline
		Fusion & \textbf{49.7} & \textbf{0.72} \\
		\hline
		\hline
		\textbf{Eval-set} & \textbf{F1} & \textbf{Error rate} \\
		\hline
		DCASE2017 baseline & 28.4  & 0.93 \\
		\hline
		Gated-CRNN-logMel & 47.50 & 0.78 \\
		\hline
		Fusion & \textbf{51.8} & \textbf{0.73} \\
		\hline
	\end{tabular}%
	\label{tab:sed}
\end{table}

\begin{figure}[t]
	\centering
	\centerline{\includegraphics[width=\columnwidth]{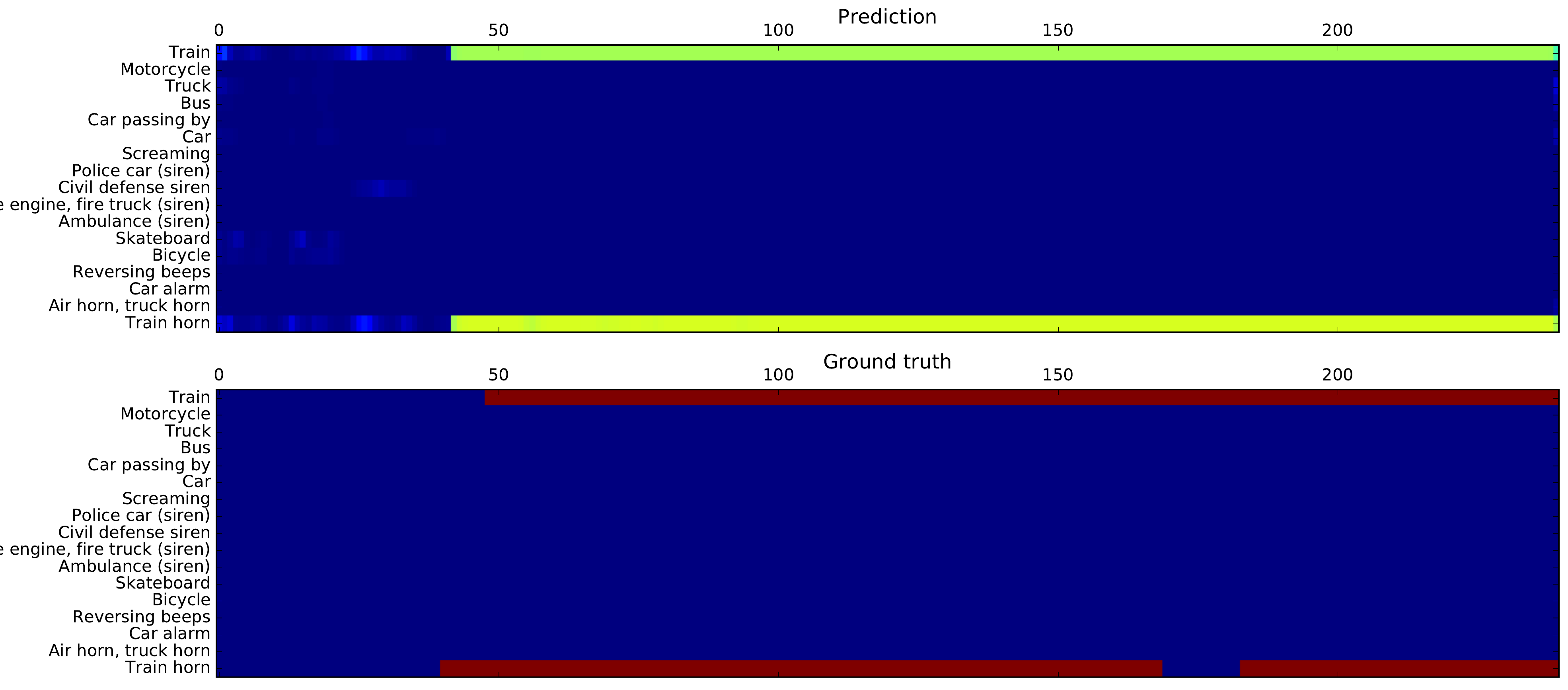}}
	\caption{An example for predicting locations along 240 frames for ``10i60V1RZkQ\_210.000\_220.000.wav'' using the proposed attention method \cite{yong2017interspeech}. The above figure is the predictions and the bottom figure is the reference.}
	\label{fig:attention_demo}
\end{figure}

Fig. \ref{fig:attention_demo} shows an example for predicting locations along 240 frames for ``10i60V1RZkQ\_210.000\_220.000.wav'' using the proposed attention method \cite{yong2017interspeech}. The above figure is the predictions and the bottom figure is the reference. It shows that the proposed method can predict the locations along the time axis.

\section{Conclusions}
This technical report briefly presents the whole system and methods for the task 4 of DCASE2017 challenge. It can get 61\% F-value for the audio tagging and 0.73 ER for the weakly supervised sound event detection subtask on the development set. Finally, we ranked first in the audio tagging sub-task on the evaluation set. We also ranked 2nd as a team in the sound event detection sub-task. For more details, please read our paper \cite{xu2017large}.

\bibliographystyle{IEEEtran}
\bibliography{refs}
%
%
%
%
%
%
%
%
%

\end{sloppy}
\end{document}